\documentclass[aps,prd,10pt,notitlepage,nofootinbib,superscriptaddress,showkeys,showpacs]{revtex4-1}

\usepackage{amsmath}
\usepackage[english]{babel}
\usepackage{graphicx,color}
\usepackage{xspace}
\usepackage{graphicx}
\usepackage{pifont,dsfont}
\usepackage{marvosym}
\usepackage{slashed}
\usepackage{subfigure}

\newcommand{\be}{\begin{equation}}
\newcommand{\ee}{\end{equation}}
\newcommand{\bqa}{\begin{eqnarray}}
\newcommand{\eqa}{\end{eqnarray}}
\newcommand{\bea}{\begin{eqnarray}}
\newcommand{\eea}{\end{eqnarray}}

\newtheorem{definition}{Definition}
\newtheorem{theorem}{Theorem}

\newtheorem{proposition}{Proposition}

\DeclareMathOperator{\tr}{tr}

\begin{document}

\title{\Large \bf New $1/N$ expansions in random tensor models}

\author{{\bf Valentin Bonzom}}\email{vbonzom@perimeterinstitute.ca}
\affiliation{Perimeter Institute for Theoretical Physics, 31 Caroline St. N, ON N2L 2Y5, Waterloo, Canada}

\date{\small\today}

\begin{abstract}
Although random tensor models were introduced twenty years ago, it is only in 2011 that Gurau proved the existence of a $1/N$ expansion. Here we show that there actually is more than a single $1/N$ expansion, depending on the dimension. These new expansions can be used to define tensor models for `rectangular' tensors (whose indices have different sizes). In the large $N$ limit, they retain more than the melonic graphs. Still, in most cases, the large $N$ limit is found to be Gaussian, and therefore extends the scope of the universality theorem for large random tensors. Nevertheless, a scaling which leads to non-Gaussian large $N$ limits, in even dimensions, is identified for the first time.
\end{abstract}

\medskip


\maketitle

\section*{Introduction}

Random tensor models were introduced twenty years ago as a generalization of random matrix models \cite{ambjorn-3d-tensors, sasakura-tensors, gross-tensors}. While a matrix is a two-dimensional array, a tensor is a $d$-dimensional array, $d\geq 2$. Partition functions of tensor models have an expansion onto Feynman graphs which are dual to triangulated $d$-dimensional pseudo-manifolds, and are therefore of interest for the path integral quantization of general relativity. They indeed realize a sum over discrete geometries and the hope is to define quantum gravity by taking the continuum limit.

However, the $1/N$ expansion of matrix models, which organizes the sum over two-dimensional topologies at large matrix size (i.e. small Newton constant) \cite{mm-review-difrancesco}, has been missing in tensor models, until Gurau eventually found one in 2011 \cite{Gur3, Gur4, GurRiv}. This $1/N$ expansion applies in particular to the set of complex tensors with measures of the form $\exp -S$ where $S$ is a combination of tensor invariants with some $U(N)$ symmetry properties \cite{1tensor}. It has triggered a full analysis of tensor models, which reproduces the key features of matrix models. The dominant contributions at large $N$ have been identified (some triangulations of the $d$-sphere) and proved to be summable \cite{critical-colored}. Their singular behavior provides a continuum limit and coupling to (non-unitary) matter generates different universality classes in the continuum whose critical exponents can be found exactly \cite{harold-hard-dimers, multicritical-dimers}.

While these developments are analogous to those obtained in matrix models, and therefore put tensor models to a somewhat similar level for the first time, the properties of the large $N$ limit are quite different. First, the multi-critical behaviors in the continuum limit correspond to the universality classes of random branched polymers (entropy exponents $\gamma=1-1/m$ in contrast to $\gamma=-1/m$ in matrix models). This was in fact expected from numerical simulations, performed in the time when no $1/N$ expansion was known in tensor models. In those numerical simulations on triangulated spheres in dimensions $d\geq 3$ (a program called Euclidean Dynamical Triangulations), it was observed that at large $N$ (small Newton constant) the partition function is dominated by stacked spheres, whose statistical properties are those of random branched polymers \cite{david-revueDT, ambjorn-scaling4D, ambjorn-revueDT}. This has been confirmed analytically using the large $N$ limit of tensor models \cite{critical-colored, harold-hard-dimers, multicritical-dimers}.

Second, random tensor models become Gaussian at large $N$ when $d\geq 3$. This is a strong universality theorem, initially proved in \cite{universality} (then in \cite{SDEs} for perturbed Gaussian measures, using field theory methods), which does not hold for random matrices. While it does not prevent critical behaviors in the continuum limit (the full large $N$ covariance crucially depends on the details of the model), this theorem directly leads to the critical exponents of random branched polymers.

The origin of the universality theorem is to be found in the structure of the dominant Feynman graphs at large $N$. It only probes the \emph{melonic} sector, where the Feynman graphs dominating the free energy maximize the number of faces (for the triangulated sphere it means maximizing the number of $(d-2)$-dimensional simplices at fixed number of top simplices), and the melonic Feynman graphs actually create \emph{more} faces than the planar ribbon graphs of matrix models (at fixed number of vertices and lines). This melonic sector has the same universality class as the stacked spheres \cite{critical-colored}, the one of random branched polymers. Moreover their combinatorics is so specific that only Gaussian behaviors can survive at large $N$.

Therefore, both the statistical properties in the continuum and the universality theorem are tied to the fact that the large $N$ limit only probes a very specific subset of triangulations. However, there is {\it a priori} more freedom in tensor models than in matrix models. The reason is that in matrix models, there exists only one invariant at order $2p$ in the matrix elements, $\tr (MM^\dagger)^{p}$, whereas many more exist when $d\geq 3$ \cite{1tensor}. (Physically, those invariants represent boundary triangulations. In 2d, they can only be loops, but as the dimension goes to three and higher, more boundary triangulations exist.) The question is thus whether other $1/N$ expansions exist in tensor models which would probe more invariants than the melonic ones.

\begin{itemize}
\item The first result of the present article is that random tensor models admit new $1/N$ expansions. They are based on the same ingredients as Gurau's, but in the large $N$ limit, they retain more than the melonic graphs.

\item However, in most cases, the large $N$ limit is still Gaussian. Although it might look surprising at first sight, the reason is that dominant contributions must have melonic subgraphs and the universality result applies to them. This is the second result, which \emph{extends the scope of the universality theorem}.

\item The third result is that in even dimension, there exists a $1/N$ expansion which is \emph{not} Gaussian at large $N$, because it is based on planar instead of melonic subgraphs. We prove it exists but do not solve this case in the present paper and leave it for future investigations.

\item Gurau's $1/N$ expansion works for tensors $T_{a_1 \dotsb a_d}$ whose indices have all the same scaling $\alpha_i N$ (where $\alpha_i$ is a constant for each index in the position $i$). It turns out that our new expansions enable to define tensor models for tensors with indices of different sizes (which do not scale together).
\end{itemize}

The article is organized as follows. In the Section \ref{sec:gurau exp} we review Gurau's $1/N$ expansion, and the solution at large $N$. The same notions are then used in the Section \ref{sec:new1/N} to make sense of new $1/N$ expansions. Their large $N$ limits are discussed in the Section \ref{sec:largeN}, including the extension of the universality theorem as well as the new large $N$ limit in even dimensions which is not Gaussian at large $N$.

\section{Gurau's $1/N$ expansion} \label{sec:gurau exp}

The presentation follows \cite{1tensor}, to which the reader is referred for more details and references.

\subsection{Invariants, colored graphs and their degree}

A tensor $T$ with components $T_{a_1\dotsb a_d}$, $a_i=1,\dotsc,N$, is a multi-dimensional array of $N^d$ complex numbers. To study large random tensors, we need a family of invariant objects constructed from its components. Let us consider the action of the external tensor product of $U(N)$,
\be \label{U(N)transfo}
T'_{a_1\dotsb a_d} = \sum_{b_1,\dotsc,b_d} U^{(1)}_{a_1 b_1} \dotsb U^{(d)}_{a_d b_d}\ T_{b_1\dotsb b_d},
\ee
for a set of $d$ independent unitary matrices $\{U^{(i)}\}$, while $\bar{T}$ transforms similarly with $\{\bar{U}^{(i)}\}$. The use of different copies of $U(N)$ acting independently on the different indices implies that invariant quantities are obtained by contracting an index of a $T$ in position $i$ with an index of a $\bar{T}$ in the \emph{same} position $i$. Taking $p$ tensors $T$ and $p$ tensors $\bar{T}$, we saturate all indices two by two, respecting their positions. Products and sums of such invariants generate the full algebra of invariants.

It is standard to represent them graphically in terms of line-colored graphs called \emph{bubbles}. To each $T$, we associate a white vertex and to each $\bar{T}$ a black vertex. The indices are represented as $d$ half-lines attached to the vertex. Because the position of the indices must be respected in the contractions, half-lines carry a label which is just the position of the corresponding index. This index is called the \emph{color} of the line and can take the values $i=1,\dotsc,d$. When an index is contracted between a $T$ and a $\bar{T}$, the two corresponding half-lines (which have the same color) are connected to form a line between a white and a black vertex.

\begin{figure}
 \includegraphics[width=8.5cm]{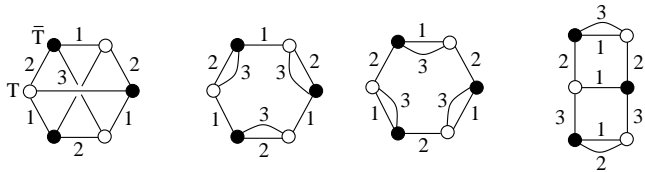}
\caption{Tensor invariants can be represented as bubbles, i.e. closed bipartite $d$-colored graphs. Here $d=3$.}
\label{fig:tensobs}
\end{figure}

When all indices are saturated, a closed, bipartite graph with vertices of degree $d$ and colored lines such that each vertex has all $d$ colors on its lines, is obtained. Examples are given in the Figure \ref{fig:tensobs}. The connected ones generate the algebra of invariants. In the remaining, bubble refers to such a bipartite, connected, $d$-colored graph, and the corresponding tensor invariant is denoted $B(T,\bar{T})$.

At $d=2$ there is a single bubble with $2p$ vertices, the loop with alternating colors 1,2, and it corresponds to the invariant $\tr(TT^\dagger)^p$.

\begin{figure}
 \includegraphics[width=4cm]{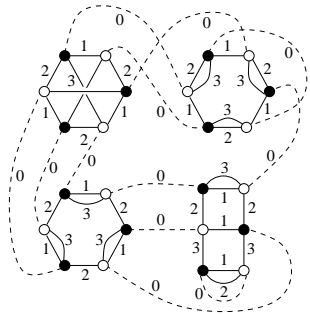}
\caption{In a Feynman graph, the vertices of the bubbles receive an additional line which connects a white vertex to a black vertex. They are represented as dashed lines with the color 0. Therefore Feynman graphs are $(d+1)$-colored graphs. Here $d=3$.}
\label{fig:tensobsgraph}
\end{figure}

Let $I$ be a finite set and $\{B_i\}_{i\in I}$ a finite set of bubbles which we use to define an invariant action
\be \label{action}
S(T,\bar{T}) = T\cdot \bar{T} +\sum_{i\in I} t_i\, B_i(T,\bar{T}),
\ee
where $\{t_i\}$ is a set of coupling constants, and $T\cdot \bar{T}$ is the natural contraction of $T$ with $\bar{T}$, corresponding to the bubble with two vertices and $d$ lines connecting them. The partition function and the free energy are
\be
Z(\{t_i\}) = \exp -F(\{t_i\}) = \int [dT\, d\bar{T}]\ \exp -N^{d-1}\,S\ .
\ee
The free energy $F$ admits an expansion onto closed connected Feynman graphs built from the bubbles $B_i$. Each vertex of a bubble $B_i$ is a $T$ or a $\bar{T}$, and therefore, due to Wick's theorem, they receive an additional line, the propagator. Each of these new lines connects a $T$ to a $\bar{T}$. By giving them the fictitious color $0$, it turns out that the connected Feynman graphs which enter the expansion of the free energy are just $(d+1)$-colored graphs, as can be seen in the Figure \ref{fig:tensobsgraph}.

The amplitude that each graph $G$ receives has a scaling with $N$ which is the key interest of this paper. The propagator is
\be
C_{a_1 \dotsb a_d, b_1 \dotsb b_d} = \frac{1}{N^{d-1}}\,\delta_{a_1 b_1}\dotsb\, \delta_{a_d b_d}.
\ee
Therefore, each line of color $0$ brings a factor $N^{-d+1}$ and the number of such lines is the half-number of vertices of $G$, denoted $p$. Each bubble brings a factor $N^{d-1}$. Moreover, the Kronecker deltas of the propagator and the contraction of indices along the colored lines of the bubbles identify the indices all along the loops with alternating colors 0 and $\mu$, for $\mu=1,\dotsc,d$. We call these loops \emph{faces with colors $0\mu$} and denote $f_{0\mu}$ their number. There is a free sum over $a_\mu=1,\dotsc,N$ for each of them, so the amplitude of a connected graph $G$ with $2p$ vertices is
\be \label{amplitude}
A(G) = N^{\sum_{\mu=1}^d f_{0\mu} - (d-1)p +(d-1)b}\ \prod_{i\in I} t_i^{b_i}.
\ee
Here $b_i$ denotes the number of bubbles $B_i$ in $G$, and $b=\sum_{i\in I}b_i$ the total number of bubbles. The key ingredient to control this scaling is a formula which counts the number of faces in terms of the number of vertices in any connected colored graph. For ribbon graphs, we know that there is a topological balance between faces and vertices (and lines) encoded in the genus of the ribbon graph \cite{mm-review-difrancesco}. Here the structure of colored graphs is probed by identifying some ribbon sub-graphs called \emph{jackets}. A jacket goes along all vertices and lines of $G$ but only a subset of faces \cite{Gur3, Gur4, GurRiv}.

\begin{definition}
 Let $\Delta\geq 3$ be an integer, $G$ be a connected, $\Delta$-colored graph with $2p$ vertices (hence $\Delta p$ lines) and $\sigma$ be a cycle on $\{1,\dotsc,\Delta\}$. We call the loops with alternating colors $\mu, \nu$, for $\mu\neq \nu$ both in $\{1,\dotsc,\Delta\}$, \emph{faces of colors $\mu,\nu$}. The jacket $J$ associated to $\sigma$ is the connected ribbon graph which contains all the faces of colors $(\sigma^q(1),\sigma^{q+1}(1))$ for $q=0,\dots,\Delta-1$ in $G$. Therefore the number of faces in $J$ is given by $f_J = 2-2g_J + \Delta p -2p$, where $g_J$ is the genus of $J$. We define the degree $\omega(G)$ of $G$ as the sum of the genera of the jackets.
\end{definition}

Notice that the degree is a positive number. Since there are $\frac{1}{2} (\Delta-1)!$ distinct jackets and every face appears in exactly $2(\Delta-2)!$ jackets, a sum over all jackets allows an over-counting of faces, leading \cite{Gur3, GurRiv, Gur4} to the following
\begin{theorem}
 For $G$ a connected $\Delta$-colored graph with $2p$ vertices and $f_G$ faces
\be \label{face counting}
f_G = \frac{(\Delta-1)(\Delta-2)}{2}p + (\Delta-1) - \frac{2}{(\Delta-2)!}\omega(G).
\ee
\end{theorem}

Therefore, for our connected $(d+1)$-colored graph $G$,
\be
\sum_{\mu=1}^d f_{0\mu} = f_G - \sum_{i\in I} b_i f_{B_i} = \frac{d(d-1)}{2}p + d - \frac{2}{(d-1)!}\omega(G) - \sum_{i\in I} b_i f_{B_i}.
\ee
When $d=2$, one simply has $f_{B_i}=1$ and the $1/N$ expansion of matrix models is reproduced straightforwardly. When $d\geq 3$, the bubbles are connected $d$-colored graphs, so that the formula \eqref{face counting} applies to the number of faces $f_{B_i}$, leading to
\be \label{face0mu counting}
\begin{aligned}
\sum_{\mu=1}^d f_{0\mu} &= \frac{d(d-1)}{2}p + d - \frac{2}{(d-1)!}\omega(G) - \sum_{i\in I} b_i\Bigl[\frac{(d-1)(d-2)}{2} p_i + d-1 - \frac{2}{(d-2)!}\omega(B_i)\Bigr],\\
&= (d-1)(p-b) + d - \frac{2}{(d-1)!}\omega(G) + \frac{2}{(d-2)!}\sum_{i\in I} b_i\,\omega(B_i).
\end{aligned}
\ee
It is shown in \cite{tree-algebra} that $\frac{2}{(d-1)!}\omega(G) - \frac{2}{(d-2)!}\sum_{i\in I} b_i\,\omega(B_i)$ is a positive integer. As a result, the free energy admits the $1/N$ expansion
\be \label{1/Nexp-free energy}
F(\{t_i\}) = N^d \sum_{\substack{\text{connected graphs $G$} \\ \text{with $d+1$ colors}}} N^{-\frac{2}{(d-1)!}\omega(G) + \frac{2}{(d-2)!}\sum_{i\in I} b_i\, \omega(B_i)}\ \frac{(-1)^b}{s(G)}\,\prod_{i\in I} (-t_i)^{b_i},
\ee
where $s(G)$ is a symmetry factor.

Similarly, the expectation values of bubbles, defined by
\be
\langle B(T,\bar{T})\rangle = e^{F}\ \int [dT\,d\bar{T}]\ B(T,\bar{T})\ \exp -N^{d-1} S(T,\bar{T}),
\ee
admit an expansion onto connected, $(d+1)$-colored graphs which contain $B$ as a marked subgraph. The scaling of each graph is obtained by assigning the scaling derived above for a $(d+1)$-colored graph and remove the contribution which comes from the insertion of $B$ as a bubble, i.e. $N^{d-1}$. Thus,
\be
\langle B(T,\bar{T})\rangle = N \sum_{\substack{\text{connected graphs $G$} \\ \text{with $d+1$ colors} \\ \text{containing $B$}}} \frac{(-1)^b}{s(G)}\quad  \prod_{i\in I} t_i^{b_i}\  N^{-\frac{2}{(d-1)!}\omega(G) + \frac{2}{(d-2)!} \omega(B) + \frac{2}{(d-2)!}\sum_{i\in I} b_i\,\omega(B_i)}
\ee

\subsection{The large $N$ limit: melonicity and universality}

For a given joint probability distribution on a random tensor, the first question is to compute the leading order expectation values of bubbles. For each bubble, there are two quantities of interest: the scaling with $N$ (the divergence/convergence degree), and its pre-factor (the amplitude). In \cite{universality}, Gurau proved the following theorem.

\begin{theorem} \label{thm:universality} A joint distribution which is invariant under the action of the external tensor product of $U(N)$, \eqref{U(N)transfo}, and which is properly uniformly bounded, converges at large $N$ to a Gaussian distribution with covariance the full 2-point function.
\end{theorem}

The criterion of being properly uniformly bounded is explained in details in \cite{universality}. Here we are only investigating the case of perturbed Gaussian measures, as defined by the action \eqref{action}, for which the proper uniform bound holds as shown in \cite{universality}. In this context, it implies that the expectation values of (connected) bubbles are bounded by $N$, independently of the bubbles,
\be
\langle B(T, \bar{T})\rangle = N^{1-\Omega(B)}\ A(B)\ U^p.
\ee
While the theorem does not provide the actual values of the scaling $\Omega(B)\geq0$, it states that the pre-factor $A(B)$ is the number of dominant Wick contractions on the bubble (the amplitude of a Gaussian model), and $U$ is the leading order, full covariance of the model ($p$ being the half number of vertices of $B$).

Remarkably, the dominant bubbles can be described explicitly and they determine $U$, as we now explain in the context of the action \eqref{action}. The scaling entering \eqref{1/Nexp-free energy} can be re-expressed as
\be
\frac{2}{(d-1)!}\omega(G) - \frac{2}{(d-2)!}\sum_{i\in I} b_i\,\omega(B_i) = \frac{2}{d!} \omega(G) + \frac{2}{d(d-2)!}\Bigl(\omega(G) - d\sum_{i\in I} b_i\,\omega(B_i)\Bigr),
\ee
where $\omega(G) - d\sum_{i\in I} b_i\,\omega(B_i)$ can be proved to be positive \cite{tree-algebra} (and proposition 2 in Appendix A of \cite{1tensor}). Therefore the scaling is a sum of positive terms, and at large $N$ only the graphs for which they vanish can survive. As a result, the free energy in the large $N$ limit is dominated by graphs $G$ with vanishing degree, built from bubbles $B_i$ whose degrees are also zero\footnote{Note that $\omega(G)=0$ implies that the degrees of its bubbles vanish.}. This is in contrast with matrix models in which all invariants $\tr(TT^\dagger)^p$ can contribute to the large $N$ limit. Here only the bubbles with $\omega(B_i)=0$ are relevant in the action.

\begin{figure}
\subfigure[An elementary melon with external color 1. It has 2 vertices and creates $\Delta(\Delta-1)/2$ new faces.]{\includegraphics[scale=0.5]{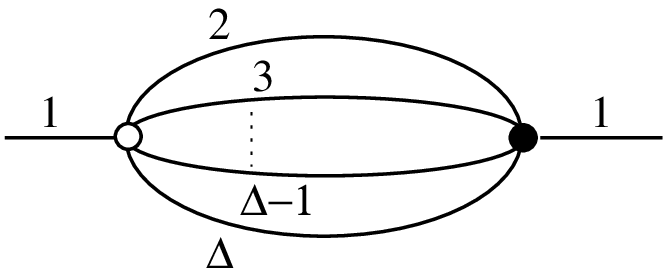} \label{fig:elementary-melon}}
\hspace{2cm}
\subfigure[A closed melonic bubble with 4 colors.]{\includegraphics[scale=0.4]{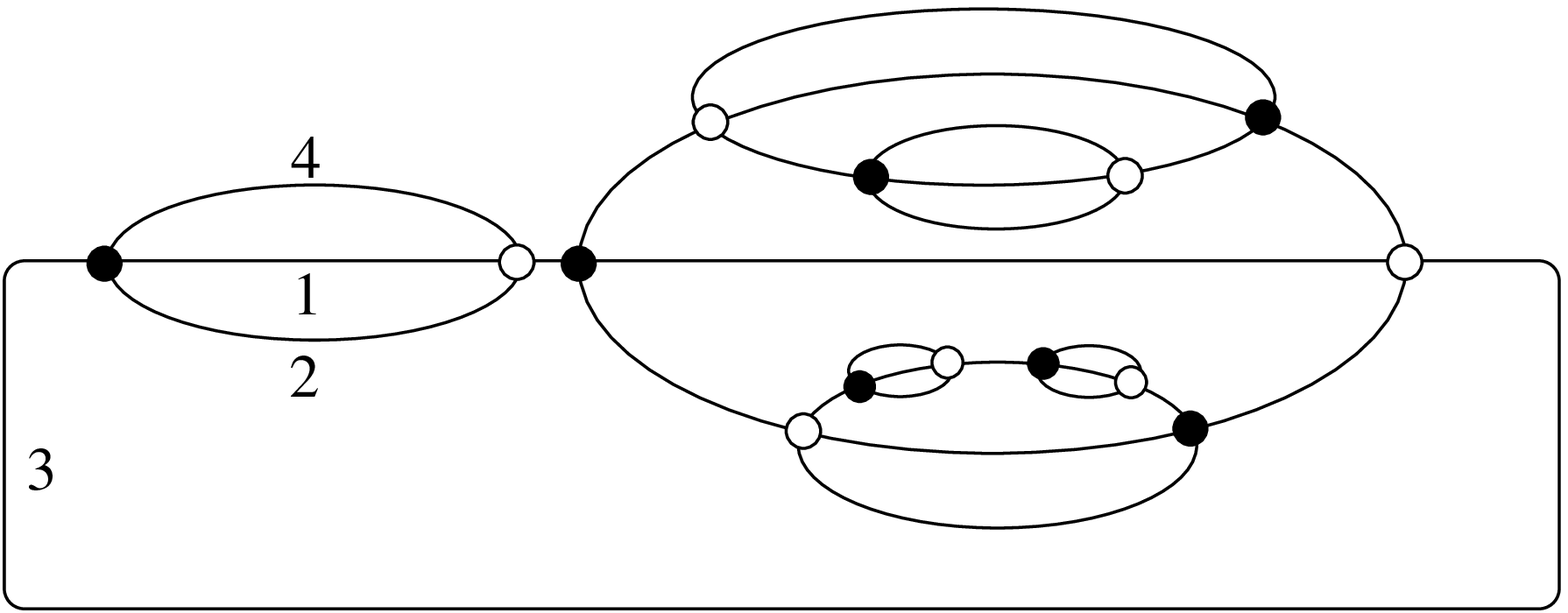} \label{fig:melon}}
\caption{Melonic colored graphs are built by recursive insertions of the elementary melon.}
\end{figure}

Bubbles and graphs of degree zero are called \emph{melonic} and built as follows. From the formula \eqref{face counting} it is found that $\omega = 0$ corresponds to maximizing the number of faces at fixed number of vertices. In addition, to keep the degree constant when adding two vertices to a graph, $\Delta(\Delta-1)/2$ faces have to be created. This can be realized locally by the 2-point insertion pictured in the Figure \ref{fig:elementary-melon}, where all lines of the vertices are paired but the two external ones (which have the same color). It is called the \emph{elementary melon}. Then it can be proved that all melonic graphs are obtained by recursive insertions of this patch on any line, starting from the closed graph with two vertices \cite{critical-colored}. An example is given in the Figure \ref{fig:melon}.

In the large $N$ limit of the model, only melonic bubbles survive and the $(d+1)$-colored graphs $G$ built from them must be melonic too. Notice that from the way melonic colored graphs are built, the vertices are inserted as \emph{canonical pairs}. To identify them within a given melonic graph, one proceeds by recursive removals of the elementary melons. The graph contains some elementary melons with obvious canonical pairs. To identify the others, one removes the elementary melons and replaces them with lines. This gives another melonic graph on which the canonical pairs of the elementary melons are trivially identified, and so on, until one is left with the graph with exactly two vertices.

\begin{figure}
\includegraphics[scale=0.9]{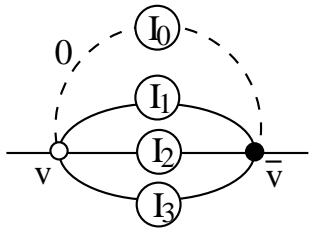}
\caption{Part of a melonic bubble (solid lines), at $d=4$, where $(v,\bar{v})$ is a canonical pair. It means that the insertions $I_1,I_2,I_3$ are melonic 2-point insertions. All melonic graphs with the additional color 0 (the propagator, drawn in dashed lines) are obtained by connecting the vertices of the canonical pair with any melonic 2-point insertion $I_0$.}
\label{fig:melonic-contraction}
\end{figure}

Given a melonic bubble $B$ with $2p$ vertices, there is only one way to build melonic $(d+1)$-colored graphs using propagators and the bubbles $\{B_i\}$ of the action, which is by connecting the vertices of the canonical pairs with lines of color 0 and inserting any melonic 2-point function on them \cite{universality, 1tensor}. The Figure \ref{fig:melonic-contraction} illustrates these melonic contractions. The full expectation value of $B(T,\bar{T})$ is thus obtained by inserting the full 2-point function $\langle T\cdot \bar{T}\rangle$ between the $p$ canonical pairs. This explains the large $N$ universal behavior,
\be
\langle B(T,\bar{T})\rangle = N\ [U(\{t_i\}]^p,\quad \text{with} \quad U(\{t_i\}) = \langle T\cdot \bar{T}\rangle/N,
\ee
for melonic bubbles with $2p$ vertices, and $\lim_{N\to\infty} \langle B\rangle/N = 0$ otherwise. To complete this solution, the completely dressed covariance $U$ has to be found as a function of the coupling constants. This is done using a Schwinger-Dyson equation, which turns into a polynomial equation \cite{1tensor},
\be \label{SDEonU}
1-U - \sum_{i\in I} t_i\,p_i\,U^{p_i} = 0,
\ee
where $p_i$ is the half-number of vertices of (melonic) $B_i$, $i\in I$.

\section{The new $1/N$ expansions} \label{sec:new1/N}

\paragraph{`Rectangular' tensors, with entries of different sizes.} Our tensor models have been so far based on polynomials which are invariant under the action of $U(N)^d$ \eqref{U(N)transfo}. However it is clear that those transformations make sense for tensors of sizes $N_1\times \dotsm \times N_d$, where $N_i$ is the range of the index in the position $i$. Similarly, the algebra of invariant polynomials is generated by the (connected) bubbles defined as before, and the tensor action \eqref{action} is still perfectly defined. The question is then how to get a large size expansion in this context.

Assume that the first $k$ indices of $T$ and $\bar{T}$ have the same size $N_1$, and the other indices, in position $k+1,\dotsc,d$ have size $N_2$. Then it would be natural to expect that Feynman graph amplitudes scale like
\be
A(G)\sim N_1^{k - \frac{2}{(k-1)!}\omega(G_1)}\ N_2^{d-k - \frac{2}{(d-k-1)!} \omega(G_2)}
\ee
where $G_1, G_2$ are respectively the sub-graphs with colors $(0,1,\dots,k)$ and $(0,k+1, \dotsc,d)$ (there is the color 0 for both because it comes from the Wick contractions and for the sake of the argument we have neglected the degrees of bubbles). The above scaling corresponds to that of two tensor models superimposed, based on two tensors (one of size $N_1^k$ and the other of size $N_2^{d-k}$).

However, it is a priori not clear how to get such scalings from a direct calculation. This is the purpose of the present article. To keep equations of reasonable lengths, we will nevertheless work with a single size $N$ for all indices. The main result will be stated for different sizes $N_i$ in the Section \ref{sec:different-sizes}, as well as the Theorem \ref{thm:gen-uni}. Generalizations of all other formulae are obvious. Our analysis will work in all cases except when there is one index whose range does not scale at all with the range of other indices. This case will be investigated elsewhere.

\paragraph{Intuitive picture.} Melonicity is a very rigid condition that we would like to partially relax. To this aim, we propose to probe the structure of colored graphs with the degrees of sub-graphs. Indeed, consider a graph $G_1$ with colors in $\{0,1,\dotsc,k\}$ and a graph $G_2$ with colors in $\{0,k+1,\dotsc,d\}$, with $k<d$ and with the same number of vertices. They can be glued together by identifying their vertices and their lines of color 0, so that we obtain a graph $G$ with colors $0,\dotsc,d$. Even when $G_1,G_2$ both have degree 0, the constructed graph $G$ may have a non-vanishing degree. It is actually possible to relate the degree of $G$ to those of its sub-graphs,
\be \label{sub-degrees}
\frac{2}{(d-1)!}\,\omega(G) = \frac{2}{(k-1)!}\,\omega(G_1) + \frac{2}{(d-k-1)!}\,\omega(G_2) +(p-b).
\ee
Here we have assumed to simplify the relation that $G_1,G_2$ are built out of bubbles with degree zero, and gluing them produces bubbles of $G$ that also have degree zero. The reason for this relation is that the degree of $G_1$ `contains' the faces with colors $(0,i)$ for $i=1,\dotsc,k$ and the degree of $G_2$ `contains' the faces with colors $(0,j)$ for $j=k+1,\dotsc,d$. Therefore, they account for all faces of colors $(0,\mu)$, as the degree of $G$ does.

However, if we re-write the large $N$ behavior \eqref{amplitude} of a Feynman graph using the degrees of sub-graphs, i.e. using \eqref{sub-degrees}, it is not uniform at fixed degrees of the sub-graphs due to the $(p-b)$ contribution in \eqref{sub-degrees}. The idea is then to absorb this contribution to the scaling by a re-definition of a coupling. Indeed, if we use a coupling $N^{d-1}/\lambda$ in front of the action, for some coupling $\lambda$, the amplitude \eqref{amplitude} becomes
\be
\begin{aligned}
A(G) &= N^{\sum_{\mu=1}^d f_{0\mu} - (d-1)p +(d-1)b}\ \lambda^{p-b}\ \prod_{i\in I} t_i^{b_i},\\
&= N^{d-\frac2{(k-1)!}\omega(G_1) - \frac2{(d-k-1)!}\omega(G_2)}\ \left(\frac{\lambda}{N}\right)^{p-b}\ \prod_{i\in I} t_i^{b_i}.
\end{aligned}
\ee
Therefore, we can reach a scaling which is uniform in the degrees of the sub-graphs with colors $0,1,\dotsc,k$ and $0,k+1,\dotsc,d$ provided the coupling $\lambda$ scales like $\lambda = N\,\kappa$ for some fixed $\kappa$. This implies that the scaling in front of the action is now $N^{d-1}/\lambda = N^{d-2}/\kappa$.

The rest of this section is devoted to making the above arguments precise and formal.

\paragraph{Color slices.} In a $(d+1)$-colored graph, the subgraphs formed by all the lines with colors in a subset of $\{0,\dotsc,d\}$ are also colored graphs and therefore have a degree. The idea of our new $1/N$ expansions is to make the Feynman graphs scale with the sum of the degrees of such colored subgraphs, instead of the degree of the full graph. To realize this, we need to chop the set of colors (index positions in the tensor) $\{1,\dotsc,d\}$ into $L$ subsets denoted $D_l$, $l=1,\dotsc,L$. Since it does not matter which color is in which subset (only the number of colors will matter), we keep the natural order on $\{1,\dotsc,d\}$ and cut that set into slices. Choose a set of integers $\{d_l\}_{l=0,\dotsc,L}$ such that $d_{l-1}<d_l$ and $d_0=0, d_L=d$. The \emph{color slice} $D_l$ is the set containing all colors between $d_{l-1}+1$ and $d_l$,
\be
D_l = \{d_{l-1}+1,\dotsc,d_l\}, \quad l=1,\dotsc,L.
\ee
It contains $\Delta_l = d_l-d_{l-1}$ colors, for $l=1,\dotsc,L$. The set of colors is partitioned into these $L$ slices,
\be
\{1,\dotsc,d\} = \bigcup_{l=1}^L D_l.
\ee
Each color slice contains at least two colors. Note that at fixed $d$, $L$ can take values between $1$ (for which $D_1$ contains all colors), and $d/2$ for $d$ even (all $D_l$ have exactly two colors), and $\lfloor d/2\rfloor +1$ for $d$ odd (all $D_l$ have exactly two colors except one with three colors).


At $d=2$ and $d=3$, this is trivial (since color singlets are not allowed): $L=1$ and there is only one subset which contains all the colors. It becomes interesting for $d\geq 4$. When $d=4$, we have two possibilities, $D_1=\{1,2,3,4\}$ with $L=1$, but also $D_1=\{1,2\}, D_2=\{3,4\}$ with $L=2$, up to color relabeling. Similarly for $d=5$, $D_1=\{1,2,3,4,5\}$ with $L=1$, and $D_1=\{1,2\}, D_2=\{3,4,5\}$ with $L=2$ up to color relabeling.

\paragraph{Classification of colored graphs.} This slicing induces a natural classification of bubbles (connected $d$-colored graphs) and Feynman graphs (connected $(d+1)$-colored graphs) according to their number of connected components in the slice $l$. For a bubble $B$, the subgraph made by all the vertices and all the lines with colors in $D_l$ is denoted $B^{(l)}$ and its number of connected components $n(B^{(l)})$. Each connected component is a connected $\Delta_l$-colored graph.

For a connected $(d+1)$-colored graph $G$ with colors $0,\dotsc,d$, $G^{(l)}$ is defined as the $(\Delta_l+1)$-colored subgraph made by all the vertices and all the lines with colors in $D_l$ plus the color 0. It has $n(G^{(l)})$ connected components.

\subsection{Introduction of scalings on the slices}

Now we choose the scaling of our tensor models as Gurau's scaling in each color slice. We know that for connected bubble invariants with $\Delta$ colors, the Feynman graphs get $\Delta+1$ colors and the correct scaling in front of the action is $N^{\Delta-1}$. The idea here is to apply this principle to all color slices $D_l$, each contributing $N^{\Delta_l-1}$, and superimpose their scalings, yielding
\be
\prod_{l=1}^L N^{\Delta_l-1} = N^{d_1-1}\times N^{d_2-d_1-1}\times \dotsb \times N^{d-d_{L-1}-1} = N^{d-L}.
\ee

Let $I$ be a finite set. We consider a finite set of connected, $d$-colored bubbles $\{B_i\}_{i\in I}$ such that they have only one connected component in the slice $l$, i.e. $n(B_i^{(l)})=1$. Remarkably we will be able to relax this requirement in the Section \ref{sec:allinvariants}, but keeping it for the time being makes the presentation simpler. The action is
\be \label{action 1connected}
S = T\cdot \bar{T} +\sum_{i\in I} t_i\,B_i(T,\bar{T}),
\ee
which has no explicit $N$-dependence. The partition function and the free energy are
\be \label{Z}
Z(\{t_i\}) = \exp -F(\{t_i\}) = \int [dT\, d\bar{T}]\ \exp -N^{d-L}\,S\ .
\ee
For $L=1$, this reduces to the scaling of the Section \ref{sec:gurau exp}.

Let us show that the above scaling leads to a $1/N$ expansion. The Feynman graphs entering the expansion of the free energy are connected $(d+1)$-colored graphs built from the bubbles $B_i$. The amplitude $A(G)$ of a graph $G$ with $2p$ vertices is found by changing $d-1$ to $d-L$ in the Equation \eqref{amplitude},
\be
A(G) = N^{\sum_{\mu=1}^d f_{0\mu} - (d-L) \ell +(d-L) b}\ \prod_{i\in I} t_i^{b_i}.
\ee
The number of faces needs to be re-expressed in terms of the number of vertices. This is done independently in each color slice. Since the bubbles $B_i$ have only one connected component $B^{(l)}_i$ in each slice and $G$ is connected, the subgraphs $G^{(l)}$ are also connected. Therefore the counting of faces in the slice $l$ is performed by applying the Equation \eqref{face0mu counting},
\be \label{faces slice l}
\sum_{\mu=d_{l-1}+1}^{d_l} f_{0\mu} = (\Delta_l -1) (p - b) + \Delta_l - \frac{2}{(\Delta_l-1)!}\omega(G^{(l)}) + \frac{2}{(\Delta_l-2)!}\sum_{i\in I} b_i\,\omega(B_i^{(l)}),
\ee
where
\be \label{positive degrees}
\frac{2}{(\Delta_l-1)!}\omega(G^{(l)}) - \frac{2}{(\Delta_l-2)!}\sum_{i\in I} b_i\, \omega(B_i^{(l)}) \geq 0.
\ee
Summing over the slices $l=1,\dotsc,L$, we get
\be
\sum_{\mu=1}^d f_{0\mu} = (d-L)(p-b)+ d - \sum_{l=1}^L \biggl[\frac{2}{(\Delta_l-1)!}\omega(G^{(l)}) + \frac{2}{(\Delta_l-2)!}\sum_{i\in I} b_i\, \omega(B_i^{(l)})\biggr].
\ee
Therefore the amplitude of a graph is bounded by $N^d$,
\be
A(G) = N^{d - \sum_{l=1}^L \bigl[\frac{2}{(\Delta_l-1)!}\omega(G^{(l)}) + \frac{2}{(\Delta_l-2)!}\sum_{i\in I} b_i\, \omega(B_i^{(l)})\bigr]}\ \prod_{i\in I} t_i^{b_i},
\ee
and the free energy has the following $1/N$ expansion
\be
F(\{t_i\}) = N^d \sum_{\substack{\text{connected graphs $G$} \\ \text{with $d+1$ colors}}} N^{-\sum_{l=1}^L \bigl[\frac{2}{(\Delta_l-1)!}\omega(G^{(l)}) + \frac{2}{(\Delta_l-2)!}\sum_{i\in I} b_i\, \omega(B_i^{(l)})\bigr]}\ \frac{(-1)^b}{s(G)}\,\prod_{i\in I} t_i^{b_i}.
\ee

\subsection{Extension to arbitrary invariants} \label{sec:allinvariants}

In the above section, only bubbles with exactly one connected component per slice $l$ were considered. We now relax this requirement and show that our new scalings allow for arbitrary bubbles. The danger is that some bubbles, like the melonic ones, create too many faces at fixed number of vertices with respect to our scaling so including them would make the free energy unbounded. However, a well-chosen, bubble-dependent, large $N$ suppression can be added in the action to make the graph amplitudes scale uniformly.

The bubbles $B_i$, $i\in I$, can now have an arbitrary number $n(B_i^{(l)})$ of connected components in the slice $l$, each being a $\Delta_l$-colored graph. The degree $\omega(B_i^{(l)})$ is the sum of their degrees. We must now introduce in the action large $N$ suppressions measured by $n(B_i^{(l)})$,
\be \label{generic action}
S(T,\bar{T}) = T\cdot \bar{T} + \sum_{i\in I} t_i\ N^{L-\sum_{l=1}^L n(B_i^{(l)})}\ B_i(T,\bar{T}).
\ee
It obviously reduces to \eqref{action 1connected} when $n(B_i^{(l)})=1$ for all $l$ and $i\in I$. The partition function still has the expression \eqref{Z}. The amplitude of a connected Feynman graph is
\be \label{graph scaling}
A(G) = N^{\sum_{\mu=1}^d f_{0\mu} - (d-L) p + db -\sum_{i\in I} b_i \sum_{l=1}^L n(B_i^{(l)})}\ \prod_{i\in I} t_i^{b_i},
\ee
The formula \eqref{faces slice l} still applies to the counting of faces in the subgraphs $G^{(l)}$, but it must be used independently for each connected component of $G^{(l)}$. Indeed, since the subgraphs $B_i^{(l)}$ can have several connected components, $G^{(l)}$ also typically has several connected components, say $n(G^{(l)})$. The latter are $(\Delta_l+1)$-colored graphs and the degree $\omega(G^{(l)})$ is the sum of their degrees. By summing the formula \eqref{faces slice l} for each connected component in $G^{(l)}$, we get
\be \label{face counting slice}
\sum_{\mu=d_{l-1}+1}^{d_l} f_{0\mu} = (\Delta_l-1) p + \Delta_l\, n(G^{(l)}) - (\Delta_l-1) \sum_{i\in I} b_i\, n(B_i^{(l)}) - \frac{2}{(\Delta_l-1)!}\omega(G^{(l)}) + \frac{2}{(\Delta_l-2)!}\sum_{i\in I} b_i\,\omega(B_i^{(l)}).
\ee
When summing over $l=1,\dotsc,L$, the contribution $(\Delta_l-1) p$ becomes $(d-L)p$ and exactly cancels the opposite term in \eqref{graph scaling}. The various degrees are positive numbers and their contribution to the scaling is bounded as can be seen by applying \eqref{positive degrees} to each connected component in $G^{(l)}$. Therefore we just have to study the terms involving the numbers of connected components $n(G^{(l)}), n(B_i^{(l)})$. The numbers of connected components $n(B_i^{(l)})$ are fixed by the choice of bubbles, but it is possible to build graphs $G$ with various $n(G^{(l)})$. Since this comes with a positive sign in \eqref{face counting slice}, graphs with many connected components can be dangerous and might lead to arbitrary high scaling. We therefore need a bound on $n(G^{(l)})$.

The fact that $G$ is connected graph can be expressed as follows. If we forget the structure of the bubbles $B_i$ used to create $G$ and simply draw them as vertices, then $G$ is a closed connected graph on these vertices, meaning that each of them has at least two lines of color 0 which connect it to others. The other lines of color 0 (there are $p$ lines of color 0 in total) connect some $T$ to some $\bar{T}$ which are in the same bubble $B_i$. Clearly we cannot form $n(B_i^{(l)})$ connected components in any $B_i$ that way, because $G$ would then not be connected. We therefore have an optimization problem with a constraint (that $G$ is connected). Let us prove the following
\begin{proposition} \label{prop:connected}
The maximal number of connected components in the closed, $(\Delta_l+1)$-colored graph $G^{(l)}$ is
\be \label{connectedmax}
n_{\rm max}(G^{(l)}) = 1+ \sum_{i\in I} b_i \bigl( n(B_i^{(l)}) -1\bigr).
\ee
\end{proposition}

{\bf Proof.} First we prove that
\be \label{ngeq}
n_{\rm max}(G^{(l)}) \geq 1+ \sum_{i\in I} b_i \bigl( n(B_i^{(l)}) -1\bigr),
\ee
by exhibiting a graph $G^{(l)}$ with precisely $1+ \sum_{i\in I} b_i( n(B_i^{(l)}) -1)$ connected components. In each $B_i^{(l)} \subset G^{(l)}$, we choose one connected component and a black and a white vertex in it. Then we form a loop going through all of them. This gives one connected component which ensures that $G$ is connected. In each bubble, we are left with exactly $n(B_i^{(l)})-1$ connected components. They all can become connected components of $G^{(l)}$ by putting the lines of color $0$ between $T$s and $\bar{T}$s in the same connected component. After summing over the bubbles, this gives $\sum_{i\in I} b_i(n(B_i^{(l)})-1)$ connected components in addition to the one which makes $G$ connected.

We now use an induction on the number of bubbles in $G$. The formula of the Proposition \ref{prop:connected} trivially holds for one bubble ($b_j=0$ except $b_i=1$ for one $i\in I$), as $n(B_i^{(l)})$ connected parts can be obtained. Let us then assume that the formula holds for any graph built from $b$ bubbles, and consider $G$ to be a graph on $b+1$ bubbles, say $b_j$ bubbles of types $B_j$, for all $j\neq i\in I$ and $b_i+1$ bubbles of the type $B_i$ for a given $i\in I$. We isolate one of the bubbles $B_i$, which we simply denote $B$, and distinguish different contributions to $n(G^{(l)})$. In the slice $l$, there are some connected components with lines of color 0 only connecting $T$s and $\bar{T}$s of $B$. Denote their number $n^{(l)}_B$. There are some connected components which do not go through $B$, say $n^{(l)}_R$ of them. Finally, some connected components are shared by $B$ and other bubbles, say $n^{(l)}_{BR}$. We have $n(G^{(l)}) = n^{(l)}_B + n^{(l)}_R + n^{(l)}_{BR}$.


Let us focus on the contributions to $n^{(l)}_{BR}$. Each of these connected parts goes along some lines of color $0$ between $B$ and other bubbles. They must actually go along an even number of such lines. Moreover each line of color 0 belongs to exactly one connected part. Let us isolate one connected part contributing to $n^{(l)}_{BR}$. It is a \emph{connected} $(\Delta_l+1)$-colored graph. Therefore, cutting the lines of color 0 which connect it to $B$ and reconnecting them in any way which disconnects the part included in $B$ from the other bubbles leads to a graph with \emph{two} connected parts. Repeating this process for all connected parts contributing to $n_{BR}^{(l)}$, all the lines of color $0$ which connect $B$ to the other bubbles are cut and we see that the number of connected components of $G^{(l)}$ is strictly less than the number of connected components which can be obtained from $B$ alone plus the number of connected components which can be obtained from the $b$ remaining bubbles. Therefore, the formula \eqref{connectedmax} can be used on the $b$ remaining bubbles to get
\be
\begin{aligned}
n(G^{(l)}) &< n(B_i^{(l)}) + 1+ b_i \bigl(n(B^{(l)}_i)-1\bigr)+\sum_{j\neq i \in I} b_j \bigl( n(B_j^{(l)}) -1\bigr),\\
&< 2 + (b_i+1) \bigl(n(B_i^{(l)}) -1\bigr) +\sum_{j\neq i\in I}b_j \bigl( n(B_j^{(l)}) -1\bigr).
\end{aligned}
\ee
This states that $n(G^{(l)})$ is less or equal to the right hand side of the formula \eqref{connectedmax} with $b_i+1$ bubbles of the type $B_i$. Combining with \eqref{ngeq} on $b_i+1$ bubbles $B_i$ concludes the induction.


It comes that
\be
\sum_{l=1}^L \Bigl[\Delta_l\, n(G^{(l)}) - (\Delta_l-1) \sum_{i\in I} b_i\, n(B_i^{(l)})\Bigr] \leq d - d\,b +\sum_{i\in I} b_i \sum_{l=1}^L n(B_i^{(l)}),
\ee
which means that the scaling of $A(G)$ with $N$ is bounded as follows
\be
\begin{aligned}
|A(G)| &= \vert \prod_{i\in I} t_i^{b_i}\vert\ N^{\sum_{l=1}^L \bigl[\Delta_l\bigl(n(G^{(l)}) - \sum_{i\in I}b_i(n(B_i^{(l)})-1)\bigr) - \frac{2}{(\Delta_l-1)!}\omega(G^{(l)}) + \frac{2}{(\Delta_l-2)!}\sum_{i\in I} b_i\,\omega(B_i^{(l)})\bigr]},\\
&\leq \vert \prod_{i\in I} t_i^{b_i}\vert\ N^{d- \sum_{l=1}^L \bigl[\frac{2}{(\Delta_l-1)!}\omega(G^{(l)}) + \frac{2}{(\Delta_l-2)!}\sum_{i\in I} b_i\,\omega(B_i^{(l)})\bigr]}.
\end{aligned}
\ee
The consequence is that the scaling we have chosen in the action restores a balance so that graphs scale uniformly (depending on the degrees of subgraphs), exactly like in the previous section when each $G^{(l)}$ has the maximal number of connected components and they are slightly suppressed if they have less.

\subsection{Scaling of observables}

The observables are the expectation values of bubble invariants,
\be
\langle B(T,\bar{T}) \rangle = e^F\ \int [dT\,d\bar{T}]\ B(T,\bar{T})\ \exp - N^{d-L}\,S(T,\bar{T}).
\ee
They admit an expansion over connected Feynman graphs which contain the bubble $B$ as a marked sub-graph. To find the scaling of the expectation value, it is sufficient to use the formula we have derived above for an arbitrary graph and remove the contribution that would come from the insertion of $B$ as a bubble of the action, i.e. $N^{-d+\sum_{l=1}^L n(B^{(l)})}$. Hence
\begin{multline}
\langle B(T,\bar{T}) \rangle = N^{\sum_{l=1}^L n(B^{(l)})} \sum_{\substack{\text{connected graphs $G$} \\ \text{with $d+1$ colors} \\ \text{containing $B$}}} \frac{(-1)^b}{s(G)}\quad  \prod_{i\in I} t_i^{b_i} \\
\times\  N^{\sum_{l=1}^L \bigl[\Delta_l\bigl(n(G^{(l)}) -1- \sum_{i\in I}b_i(n(B_i^{(l)})-1)\bigr) -\frac{2}{(\Delta_l-1)!}\omega(G^{(l)}) + \frac{2}{(\Delta_l-2)!} \omega(B^{(l)}) + \frac{2}{(\Delta_l-2)!}\sum_{i\in I} b_i\,\omega(B_i^{(l)})\bigr]}
\end{multline}
Notice that in the second line the total exponent of $N$ is negative or at most zero, due to \eqref{connectedmax}, so that at most $\langle B\rangle \sim N^{\sum_{l=1}^L n(B^{(l)})}$. In particular, for the full 2-point function (corresponding to the bubble with exactly two vertices), $\langle T\cdot \bar{T}\rangle \sim N^L$.

\subsection{$1/N$ expansions for `rectangular' tensors} \label{sec:different-sizes}

In this section, we generalize (without explicit proofs) the above results to the case where the indices whose positions correspond to colors in the slice $D_l$ have range $a_i = 1,\dotsc, N_l$, for $i=d_{l-1}+1,\dotsc,d_l$. The integers $N_l$, $l=1,\dotsc,L$, are different and may not scale the same.

We take the action
\be
S(T,\bar{T}) = T\cdot \bar{T} + \sum_{i\in I} t_i\ \left[\prod_{l=1}^L N_l^{1- n(B_i^{(l)})}\right]\ B_i(T,\bar{T}),
\ee
and the generalization of the free energy is
\be
\exp -F(\{t_i\}) = \int [dT\, d\bar{T}]\ \exp - \left[\prod_{l=1}^L N_l^{\Delta_l-1}\right]\,S\ .
\ee
Then the amplitude of a Feynman graph reads
\be
A(G) = \prod_{l=1}^L N_l^{\sum_{\mu=d_{l-1}+1}^{d_l} f_{0\mu} - (\Delta_l-1)(p-b) - \sum_{i\in I} b_i n(B_i^{(l)})}\ \prod_{i\in I} t_i^{b_i},
\ee
which rewrites as
\be
A(G) = \prod_{l=1}^L N_l^{\Delta_l\bigl(n(G^{(l)}) - \sum_{i\in I}b_i(n(B_i^{(l)})-1)\bigr) - \frac{2}{(\Delta_l-1)!}\omega(G^{(l)}) + \frac{2}{(\Delta_l-2)!}\sum_{i\in I} b_i\,\omega(B_i^{(l)})}\ \prod_{i\in I} t_i^{b_i}.
\ee
The free energy obviously expands onto connected $(d+1)$-colored graphs and is bounded by $\prod_{l=1}^L N_l^{\Delta_l}$, meaning that there is a well-defined $1/N_l$ expansion for all $l=1,\dotsc,L$.

\section{The large $N$ limits} \label{sec:largeN}

In the large $N$ limit, the free energy is dominated by graphs $G$ whose subgraphs $G^{(l)}$ have vanishing degrees $\omega(G^{(l)}) =0$ for all $l=1,\dotsc,L$. There are two cases to be distinguished.
\begin{itemize}
 \item For a slice of two colors, $\Delta_l=2$, $G^{(l)}$ is a 3-colored graph and its degree $\omega(G^{(l)})$ is the sum of the genera of each connected components. Therefore, in the large $N$ limit, the connected components of $G^{(l)}$ must be planar \cite{mm-review-difrancesco}. That gives no restriction on the bubbles $\{B_i\}$, except that there must exist graphs $G$ whose subgraphs $G^{(l)}$ for all $l$ are planar at the same time.
 \item For slices with at least three colors, i.e. $\Delta_l\geq 3$, the connected components of $G^{(l)}$ must be $(\Delta_l+1)$-colored \emph{melonic} graphs. This can only be true when $\omega(B_i^{(l)}) =0$, i.e. the bubbles must be melonic on the colors $d_{l-1}+1,\dotsc,d_l$.
\end{itemize}

This framework therefore allows more bubbles $B_i$ to contribute in the large $N$ limit than the scaling of the Section \ref{sec:gurau exp} where only melonic bubbles on the colors 1 to $d$ are relevant. For example, at $d=4$, with $L=2$ and with the color decomposition $D_1 = \{1,2\}, D_2=\{3,4\}$, any bubble can be added to the action and will {\it a priori} contribute non-trivially, as far as the sub-graphs $G^{(l)}$ can be planar at the same time. At $d=5$ with $L=2$, and $D_1 = \{1,2\}, D_2=\{3,4,5\}$, $G^{(1)}$ must be planar and $G^{(2)}$ melonic. This implies that all bubbles which are melonic on the colors $3,4,5$ can {\it a priori} contribute (see below why some may not) and in particular the bubbles which are melonic on the five colors always contribute.

\subsection{Extension of the Universality theorem}

Since the large $N$ scaling of a bubble grows like $N^{\sum_l n(B^{(l)})}$, the measure defined by the exponential of the action \eqref{generic action} is not properly uniformly bounded so that the theorem \ref{thm:universality} cannot apply, at least directly. In fact, up to this global scaling $N^{\sum_l n(B^{(l)})}$, the graph amplitudes which contribute to an expectation value scales with the degrees of the subgraphs\footnote{The reason for this is precisely that the scaling $N^{\sum_l n(B_i^{(l)})}$ which would affect the amplitudes every time $B_i$ is inserted has been corrected in the action \eqref{generic action}.}, which means that the universality theorem actually applies in all color slices. This leads to the following theorem (that we write for tensors with different sizes).

\begin{theorem} \label{thm:gen-uni} Let $T$ be a random tensor of size $N_1^{\Delta_1}\times \dotsm\times N_L^{\Delta_L}$, with $\Delta_l\geq 3$ for all $l=1,\dotsc,L$, whose joint distribution is invariant under $U(N_1)^{\Delta_1} \times \dotsm\times U(N_L)^{\Delta_L}$ and properly uniformly bounded on each color slice $1,\dotsc,L$. When $N_1,\dotsc,N_L$ go to infinity (at a priori uncorrelated speeds), the joint distribution converges to a Gaussian distribution with covariance the full 2-point function,
\be
\langle B(T,\bar{T})\rangle = \prod_{l=1}^L N_l^{n(B^{(l)}) - \Omega_l(B)}\ A(B)\ U^p,
\ee
where $\Omega_l(B)\geq 0$ for all $l$, $A(B)$ is the number of dominant Wick contractions and $U = \langle T\cdot \bar{T}\rangle/\prod_{l=1}^L N_l$.
\end{theorem}

The proof just repeats that of \cite{universality} on each color slice. The additional difficulty is to control the number of connected components in each subgraph $G^{(l)}$, which can be done using the Proposition \ref{prop:connected}.

To get $U$ in our models we need the expectation values of the bubbles which are relevant in the action at large $N$, i.e. those which exactly scale like $N^{\sum_l n(B^{(l)})}$ (again, we only discuss the case $N_l=N$). As already explained, they are the ones whose sub-bubbles $B_i^{(l)}$ are melonic.

The key point of the universality theorem for melonic bubbles is that they admit only one way to add propagators and bubbles from the set $\{B_i\}$ so as to get melonic graphs. Vertices in a melonic bubble come in pairs $(v,\bar{v})$, such that the lines connecting $v$ to $\bar{v}$ carry melonic 2-point insertions. The graphs contributing to the expectation value of the bubble have one additional color (the color 0) and melonic contributions are all such that the lines of color 0 connect the vertices $v$ and $\bar{v}$ of the canonical pairs with arbitrary 2-point (melonic) insertions, as shown in the Figure \ref{fig:melonic-contraction}. Therefore, the full contribution is obtained by inserting the full 2-point function of the model between each canonical pair of vertices of the bubble. 

\begin{figure}
\includegraphics[scale=0.65]{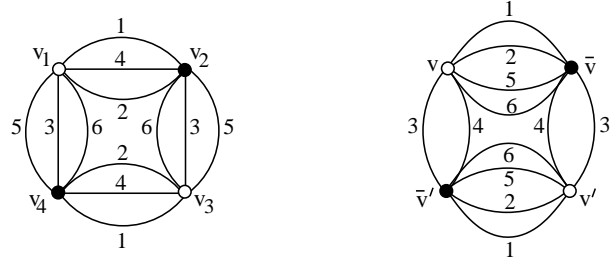}
\caption{Here $d=6$. These two bubbles are melonic on the colors 1,2,3 and on the colors 4,5,6. However, on the left, the canonical pairs of vertices for the colors 1,2,3 are $(v_1,v_2)$ and $(v_3,v_4)$, while there are $(v_1,v_4)$ and $(v_2,v_3)$ for the colors 4,5,6. Therefore there is no way to add the color 0 while preserving melonicity on the colors 1,2,3 and on the colors 4,5,6 at the same time, which makes this bubble irrelevant at large $N$. On the right of the figure, the canonical pairs coincide for the subgraphs with colors 1,2,3 and 4,5,6 (there are $(v,\bar{v})$ and $(v',\bar{v}')$), so that this bubble is relevant and has a non-vanishing expectation value at leading order.}
\label{fig:non-melonic-ex}
\end{figure}

This reasoning holds in our new models on the color slice $l$ when the slice has at least three colors. All graphs $G^{(l)}$ contributing to the expectation value of a bubble $B$ must connect the vertices of the canonical pairs of $B^{(l)}$. And it must be so for all slices $l=1,\dotsc,L$. This requires that the canonical pairs of vertices are the same for all subgraphs $B^{(l)}$ in the slices with $\Delta_l\geq 3$. If this does not hold, the large $N$ limit of the expectation value is zero, and this bubble is irrelevant in the action. Examples are given in the Figure \ref{fig:non-melonic-ex}.

\begin{figure}
 \includegraphics[scale=0.7]{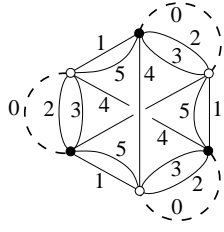}
 \caption{\label{fig:melonic-nonplanar} This is a Gaussian contraction (3 dashed lines of color 0) on a 5-colored bubble (with solid lines). The bubble is melonic on the colors 1,2,3 and the contraction is melonic on the colors 0,1,2,3. However, it is not planar on the colors 0,4,5. To see it, we count the number of faces: $f_{04}=f_{05}=f_{45} =1$, hence the genus is $g=1$. The expectation value of this bubble has a vanishing contribution at leading order in the Gaussian ensemble with $D_1=\{1,2,3\}, D_2=\{4,5\}$.}
\end{figure}

If there are in addition some slices with $\Delta_l=2$, it is also necessary to check that the single melonic contraction yields a planar contribution for $G^{(l)}$. It is not always the case, as shown in the Figure \ref{fig:melonic-nonplanar}.

Therefore, the leading order expectation value $\langle B\rangle \sim N^{\sum_{l=1}^L n(B^{(l)})}$ is either zero or Gaussian with pre-factor 1. Assuming $B$ is a connected bubble with $2p$ vertices with non-vanishing leading order,
\be \label{factorization}
\langle B(T,\bar{T}) \rangle = N^{\sum_{l=1}^L n(B^{(l)})}\ \bigl[U(\{t_i\})\bigr]^p,
\ee
where $U = \langle T\cdot \bar{T}\rangle/N^L$ is the full 2-point function, which depends on the coupling constants $\{t_i\}$. For instance at $d=5$, $L=2$, $D_1=\{1,2,3\}, D_2=\{4,5\}$, one has
\be
\left\langle \begin{array}{c} \includegraphics[scale=0.4]{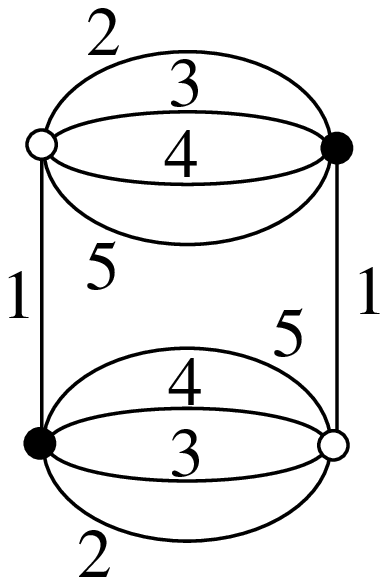} \end{array} \right\rangle = \begin{array}{c} \includegraphics[scale=0.4]{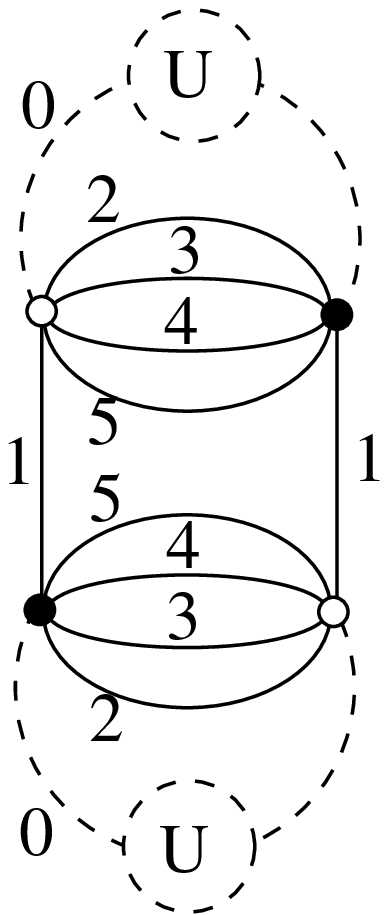} \end{array} = N^3\,U^2,\qquad \left\langle \begin{array}{c} \includegraphics[scale=0.4]{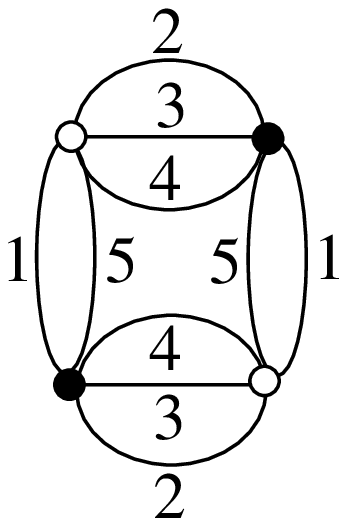} \end{array} \right\rangle = \begin{array}{c} \includegraphics[scale=0.4]{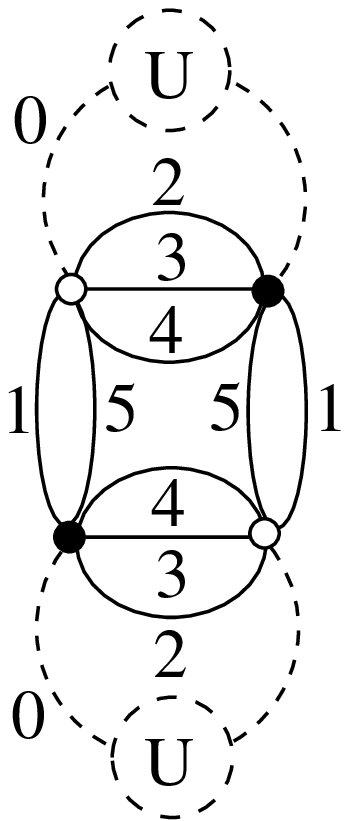} \end{array} = N^2\,U^2,
\ee
which explains why, after re-scaling the bubble on the left by $1/N$, these two bubbles can equally contribute to the large $N$ free energy, while with the usual scaling, $L=1$, the second bubble would not (it is not melonic on the five colors).

The full 2-point function $U(\{t_i\})$ is obtained explicitly thanks to a Schwinger-Dyson equation. We start with the trivial identity
\be
\sum_{a_1,\dotsc,a_d} \int [dT\,d\bar{T}]\ \frac{\partial}{\partial T_{a_1 \dotsb a_d}}\Bigl( T_{a_1 \dotsb a_d}\ \exp -N^{d-L}\,S(T,\bar{T})\Bigr) = 0.
\ee
Then we write explicitly the action of the derivative, which yields
\be
N^d - N^{d-L}\,\langle T\cdot \bar{T}\rangle - \sum_{i\in I} t_i\,p_i\,N^{d-\sum_{l=1}^L n(B_i^{(l)})}\,\langle B_i(T,\bar{T})\rangle = 0.
\ee
Notice that all terms have the same scaling $N^d$. Using the factorization \eqref{factorization} property, a polynomial equation on $U$ is found
\be
1 - U - \sum_{i\in I} t_i\,p_i\,U^{p_i} = 0,
\ee
which is the same equation as \eqref{SDEonU}. This may be surprising because the bubbles $B_i$ are not necessarily melonic on their $d$ colors. However, with our scaling, the factorization property \eqref{factorization} holds as soon as one subgraph $B_i^{(l)}$ is melonic.

The result of this section can also be derived by a detailed investigation of the full set of Schwinger-Dyson equations, following the method of \cite{SDEs}. We have checked it explicitly, but do not reproduce it as it is quite long.

\subsection{Non-Gaussian models at large $N$}

We have been able to extend the scope of the universality theorem for large random tensors when a color slice $\Delta_l$ contains three or more colors. But when $d$ is even, it is possible to use a scaling such that all color slices have exactly two colors, in which case the theorem does not apply. While we have not solve these models at the end of the day, we can prove here that
\begin{itemize}
\item this class of models contain all the possible behaviors of random matrix models,
\item there exist \emph{non}-Gaussian models at large $N$ which cannot be written as matrix models. They are the first examples of genuine random tensor models which do not satisfy the universality.
\end{itemize}
These results are rooted in the fact that for color slices with exactly two colors (and only in this case), the large $N$ scaling is controlled by the genera of ribbon sub-graphs instead of the degree of sub-graphs with four or more colors. In particular, the large $N$ limit requires sub-graphs to be planar instead of melonic and it is well-known that the behavior of planar ribbon graphs is very different from that of melonic graphs (and leads to non-Gaussian large $N$ limit).

Let $d$ be even, $L=d/2$ and $\Delta_l=\{2l-1,2l\}$ for $l=1,\dotsc,d/2$. For a bubble $B$, $n(B^{(l)})$ is the number of loops with alternating colors $2l-1,2l$ in $B$. The free energy $F$ is given by
\be
\exp -F = \int [dT\,d\bar{T}]\ \exp -N^{d/2}\,\Bigl(T\cdot \bar{T} + \sum_{i\in I} t_i\,N^{d/2-\sum_{l=1}^{d/2} n(B_i^{(l)})}\,B_i(T,\bar{T})\Bigr).
\ee
If $B_i$ has a single loop in each color slice, then it receives no extra scaling in the action but the one in front of the action. Let us focus on that case to simplify a bit the scalings. The Feynman graphs of that model are $(d+1)$-colored graphs and the subgraphs $G^{(l)}$ are 3-colored graphs, with colors $(0,2l-1,2l)$. Their degrees are simply the genera $g^{(l)}(G) = (2-f(G^{(l)}) +p)/2$, where $f(G^{(l)})$ is the number of faces of $G^{(l)}$. The amplitude of a connected Feynman graph in that model is
\be
A(G) = N^{\sum_{l=1}^{d/2} 2-2g^{(l)}(G)}\ \prod_{i\in I} t_i^{b_i}.
\ee
In other words, one has the scaling of a matrix model for each subgraphs $G^{(l)}$. This implies that at large $N$, there is no melonicity involved anymore, but only planar graphs. Since it is well-known that planar graphs typically lead to non-Gaussian large $N$ behaviors, this class of models looks promising to escape the universality of large random tensors.

\subsubsection{A matrix model for 4-dimensional spheres at large $N$}

\begin{figure}
\includegraphics[scale=0.75]{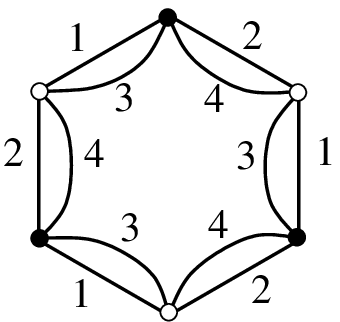}
\caption{\label{fig:MMdaggercube} This bubble can be recast as a matrix invariant, $\tr (MM^\dagger)^3$ (The non-Gaussian behavior is thus well-known).}
\end{figure}

\begin{figure}
\includegraphics[scale=0.75]{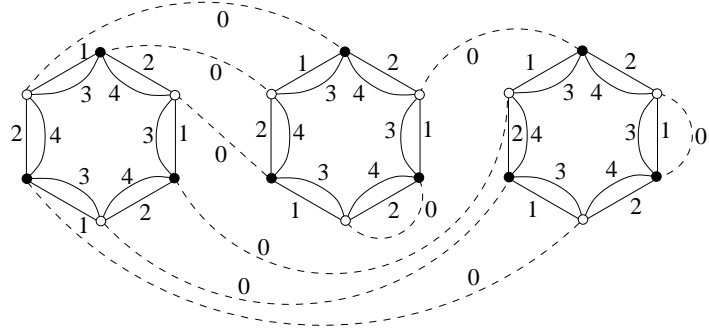}
\caption{\label{fig:4-sphere} This is a colored graph appearing in the model \eqref{MMdaggercube} and dual to a triangulation of the 4-sphere (because the jacket associated to the cycle $(01342)$ is planar).}
\end{figure}

We set $d=4$. A simple example of bubble is the one in the Figure \ref{fig:MMdaggercube}. Notice however that it can be recast as a matrix invariant of order 6 for a matrix of size $N^2\times N^2$. The matrix is $M_{AB} = T_{a_1a_2a_3a_4}$ with $A=(a_1, a_3)$ and $B=(a_2,a_4)$ and the invariant is then $\tr (M M^\dagger)^3$. Then our tensor model is just a matrix model, which can be solved using matrix model techniques. Although it is not a new result {\it per se}, it is remarkable that such a matrix model (and actually all matrix models for a generic complex matrix of size $N^2\times N^2$ and a polynomial potential $\sum_i t_i\,\tr (MM^\dagger)^i$ in the action) produces only 4-spheres in the large $N$ limit.
\be \label{MMdaggercube}
-\ln \int [dT\,d\bar{T}]\ \exp -N^2 \left(T\cdot\bar{T} + \begin{array}{c} \includegraphics[scale=.6]{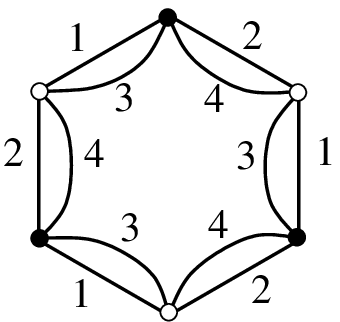}\end{array}\right)
\underset{\text{{\footnotesize large $N$}}}{=} \sum_{\substack{\text{Triangulations}\\ \text{4-sphere}}} N^4\,\frac{(-1)^b}{s(G)}.
\ee
The set of triangulations is one-to-one to the set of planar maps obtained by the corresponding matrix model at large $N$. An example is given in the Figure \ref{fig:4-sphere}.

\subsubsection{Genuine tensor models which are non-Gaussian at large $N$}

\begin{figure}
\includegraphics[scale=.75]{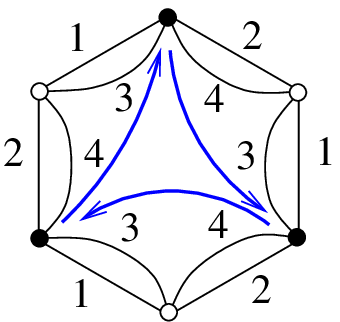}
\hspace{2cm}
\includegraphics[scale=0.75]{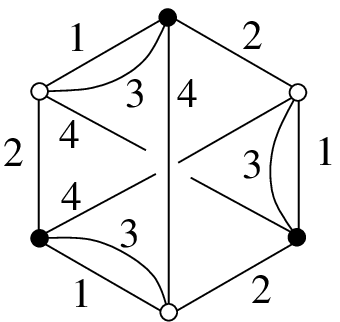}
\caption{\label{fig:MMdaggercubetwisted} This is a genuine tensor invariant because the loop of colors 34 is twisted with respect to the loop with colors 12.}
\end{figure}

We now show that similar behaviors can also be produced from genuine tensor model (where the observables cannot be written as matrix traces). Consider the bubble depicted in the Figure \ref{fig:MMdaggercubetwisted}. It is obtained by first drawing the loop with colors $1,2$, and then the loop with colors $3,4$ while applying a cyclic permutation on the three black vertices, so as to produce a twist in the loop.

\begin{figure}
\begin{center}
\subfigure[\ $f_{03}=3$, $f_{04}=1, g_{034}=0$]{\includegraphics[scale=0.65]{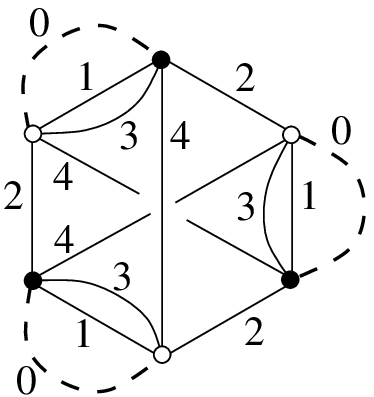} \label{fig:gaussian-contraction1}}
\subfigure[\ $f_{03}=2$, $f_{04}=2, g_{034}=0$]{\includegraphics[scale=0.65]{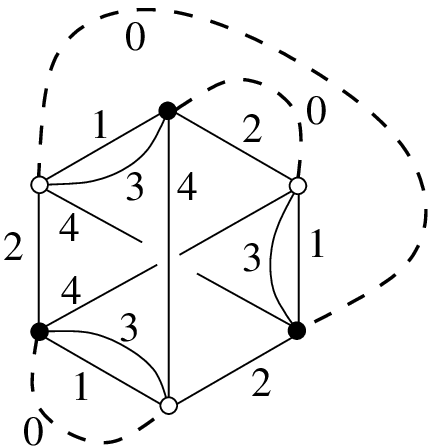} \label{fig:gaussian-contraction2}}
\subfigure[\ $f_{03}=2$, $f_{04}=2, g_{034}=0$]{\includegraphics[scale=0.65]{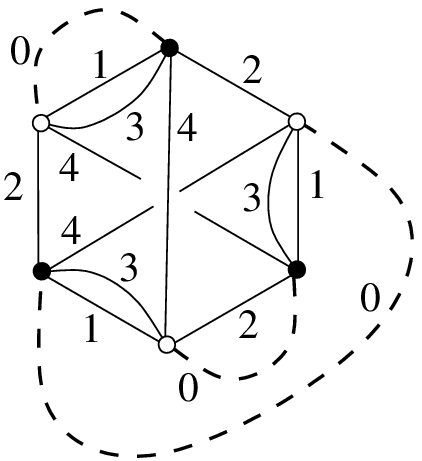} \label{fig:gaussian-contraction3}}
\subfigure[\ $f_{03}=2$, $f_{04}=2, g_{034}=0$]{\includegraphics[scale=0.65]{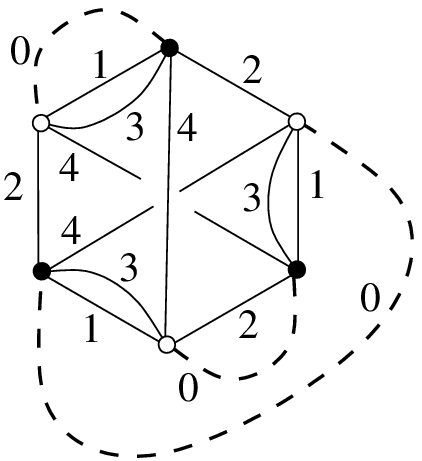} \label{fig:gaussian-contraction4}}
\subfigure[\ $f_{03}=1$, $f_{04}=1, g_{034}=1$]{\includegraphics[scale=0.65]{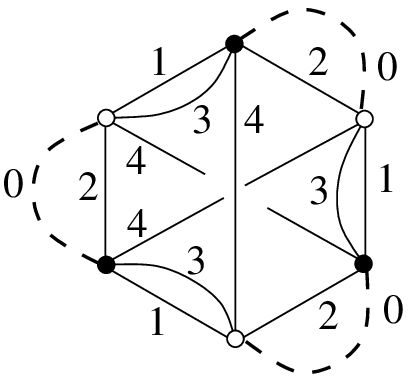} \label{fig:gaussian-contraction5}}
\end{center}
\caption{\label{fig:gaussianplanarcase} The graphs represent Gaussian contractions of a bubble with propagators (dashed lines of color 0). The sub-graphs with colors 012 are always planar. Below each graph we write the number of faces of colors 03 and 04. The genus of the sub-graphs with colors 034 are found by $g_{034} = (2-f_{03}-f_{04}-1+3)/2$. It shows that the contraction on the far right is \emph{not} planar on the colors 034.}
\end{figure}

\begin{figure}
\includegraphics[scale=0.7]{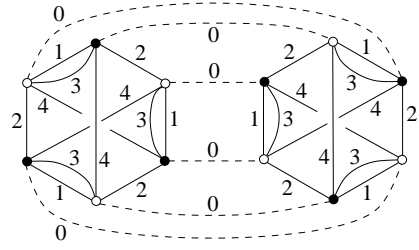}
\caption{This is a two bubble graph which is planar on the colors 012 and 034, and therefore scales like the Gaussian planar contractions in the Figure \ref{fig:gaussianplanarcase}. But it is a non-Gaussian contribution to the free energy. \label{fig:nongaussian}}
\end{figure}

A Gaussian contraction which is planar on the colors 0,1,2 and planar on 0,3,4 is shown in the Figure \ref{fig:gaussian-contraction1}, which therefore scales like $N^4$. Moreover, there exist non-Gaussian contributions to the large $N$ limit, as shown in the Figure \ref{fig:nongaussian}. It is made of two bubbles, and cutting any two lines of color 0 clearly does not disconnect the graph (this is a 6-point contribution, hence non-Gaussian). It has 6 lines of color 0 and 12 faces with colors $(0i)$ for $i=1,2,3,4$, which leads to the scaling $N^{2\times 2- 6\times 2 +12} = N^4$, as required.

Moreover, we insist on the following point. The dominant graphs in the models presented in this section must have their subgraphs with colors 0,1,2 planar, with colors 0,3,4 planar and so on. In the model built with the bubble from the Figure \ref{fig:MMdaggercube}, the subgraphs with colors 0,3,4 are planar as soon as as the subgraphs with colors 0,1,2 are planar (because they simply are the same). In particular, the expectation value of this bubble in the Gaussian ensemble of covariance $\lambda$ is the same (after rescaling by $N^{-2}$) as the expectation value of $\tr (MM^\dagger)^3$ (rescaled by $N^{-1}$), i.e. $\lambda^3\times 5$, where $5$ is the Catalan number $c_3$ counting the number of planar contractions \cite{mm-review-difrancesco}. However, with the bubble of the Figure \ref{fig:MMdaggercubetwisted}, the subgraphs with colors 0,1,2 and with colors 0,3,4 are different and may not be planar at the same time. For instance, in the Gaussian ensemble of covariance $\lambda$, the expectation value of this bubble (rescaled by $N^{-2}$) is $\lambda^3\times 4$. The Figures \ref{fig:gaussian-contraction1}, \ref{fig:gaussian-contraction2}, \ref{fig:gaussian-contraction3}, \ref{fig:gaussian-contraction4}, \ref{fig:gaussian-contraction5} indeed shows that only four of the five planar contractions for the subgraph with colors 0,1,2 are also planar for the subgraph with colors 0,3,4.

\begin{figure}
\includegraphics[scale=.75]{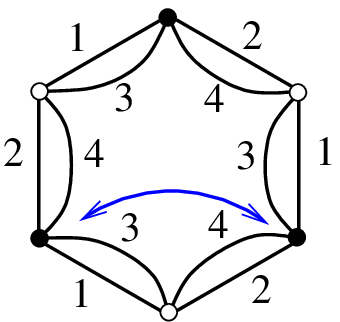}
\hspace{3cm}
\includegraphics[scale=0.75]{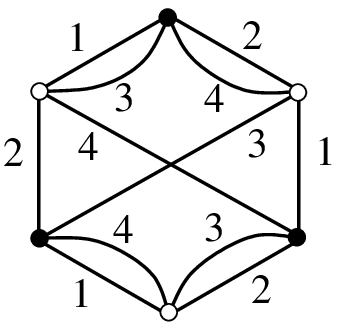}
\caption{\label{fig:MMdaggercube-transpose} On the left is a matrix invariant. To get a genuine tensor invariant, we exchange the two black vertices on the loop with colors 3, 4, which gives the bubble on the right.}
\end{figure}

\begin{figure}
\begin{center}
\subfigure[A set of Wick contractions on the bubble of Figure \ref{fig:MMdaggercube-transpose} provides a Gaussian contribution]{\includegraphics[scale=0.75]{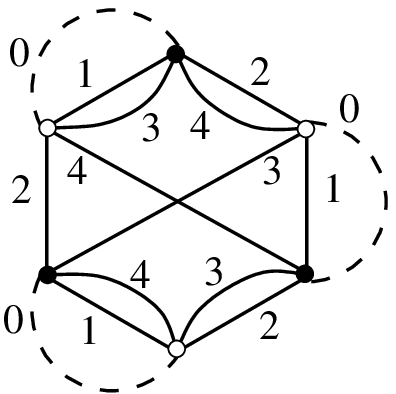}
\label{fig:transpose-gaussian}}
\hspace{1cm}
\subfigure[This is a six point contribution since six lines of color 0 need to be cut to disconnect the graph (the two bubbles are complex conjugate).]{\includegraphics[scale=0.75]{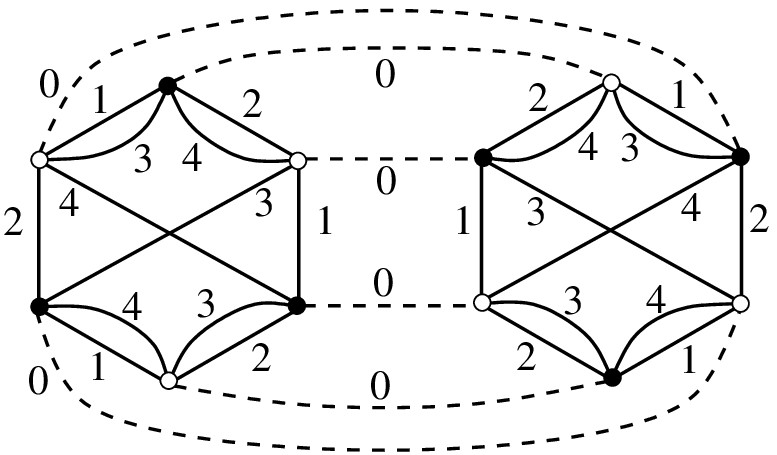}
\label{fig:nongaussian-transpose} }
\end{center}
\caption{The dashed lines of color 0 represent propagators. On the left is a Gaussian contribution and on the right a non-Gaussian contribution.}
\end{figure}

Another example of a bubble that leads to non-Gaussian large $N$ behavior is given in the Figure \ref{fig:MMdaggercube-transpose}. It is obtained by first drawing the loop with colors $1,2$, and then the loop with colors $3,4$ while exchanging two black vertices, producing a new twist in the loop. The Figure \ref{fig:transpose-gaussian}  displays Gaussian contractions which contribute to the leading order. The Figure \ref{fig:nongaussian-transpose} is then a 6-point contribution (the two bubbles are complex conjugate), which scales like $N^4$ as expected.

More generally, the set of bubbles with a single loop of colors $(1,2)$ and a single loop with colors $(3,4)$ appears as a natural generalization of the matrix trace invariants. Compared with other bubbles (in particular melonic ones), the diagrammatic coming from these observables is certainly much closer to matrix models, and will be the subject of future prospects.

\section*{Conclusion}

In this paper, we have shown that defining $1/N$ expansions for `rectangular' tensors leads to generalizing Gurau's $1/N$ expansion to a family of well-defined expansions. There are more available expansions as the dimension $d$ increases (the new scalings we are proposing appear at $d=4$). The key idea is to slice the set of indices of the random tensor into groups of two or more indices (the indices of different can have different ranges) and associate to a colored graph the degrees of its sub-graphs determined by the slicing, as described in the Section \ref{sec:new1/N}. This re-organizes Gurau's expansion, enhancing the contribution of previously suppressed graphs, in such a way that for instance more than the melonic sector contributes at large $N$. However, as soon as there is a slice with at least three colors, the large $N$ behavior can only be Gaussian, because the corresponding sub-graphs must be melonic and the universality theorem applies to them. This result is an unexpected extension of the universality theorem of \cite{universality}.

One motivation of this work was to identify scalings leading to non-Gaussian large $N$ limits, thereby stepping out of the universality. This is indeed possible for even $d$, by looking at the genus of ribbon sub-graphs carrying only some subsets of colors. Therefore, it seems that such a scaling is closer to matrix models than to the large $N$ limit of random tensors known so far. The next step is thus to find the large $N$ solution (and eventually the solution at all orders), describe its continuum limits and critical exponents. While we expect them to be different of the universality classes already found in tensor models, it is not clear yet whether it will coincide with the critical behaviors of matrix models or be completely new.

A question that has not been discussed in the main text of the paper is the topology of the graphs. Indeed, the quantum gravity motivation in random tensors is that the bubbles are dual to $(d-1)$-dimensional pseudo-manifolds and the Feynman graphs in the expansion of the free energy are dual to $d$-dimensional pseudo-manifolds. It is known that the melonic graphs are dual to triangulated spheres, generalizing the relation between planar graphs and discretizations of the sphere in two dimensions. However it is also known that not all colored triangulations of the sphere are melonic. Quite evidently, our new large $N$ limits pick up contributions from spheres which are not melonic, which is necessary to get more realistic quantum gravity models. However, for the new scalings in even dimensions which lead to non-Gaussian large $N$ behaviors, non-spherical topologies typically show up at leading order. This suggests that getting a tensor model which is not melonic and not Gaussian but generates only spheres at large $N$ is a very non-trivial problem and at this stage, it seems that a double scaling limit could be the best way.

Finally, we would like to mention a more speculative idea. It has been so far observed that although the degree reduces to the genus at $d=2$, the combinatorics of the dominant graphs is quite different between matrix and tensors for $d\geq 3$. The physical consequences are the universality theorem for $d\geq 3$, different statistical behaviors and hence different critical exponents in the continuum limit. However, the present paper shows that tensor models admit matrix-like scalings (i.e. with the genus of ribbon sub-graphs). Therefore, it is natural to ask whether interpolating between matrix-like and tensor-like behaviors is possible. While appealing, this idea seems difficult to realize. It would certainly necessitate some analytical continuations on $d$, $L$ (the number of color slices), and on the number of colors in the slices. The difficulty is that the difference between $d=2$ and $d\geq 3$ only appears in the formula for the degree \eqref{face counting} and it involves faces, i.e. closed subgraphs labeled by pairs of colors. It is unclear how to go to non-integer numbers of colors from that view.

If that question can be addressed, we think it could be through the Schwinger-Dyson equations. Indeed, the set of Schwinger-Dyson equations found in \cite{bubble-algebra} is formally the same as for the new scalings. They involve the same set of operations on bubbles. Only the powers of $N$ are modified so as to change the balance between the different orders in the $1/N$ expansion of the bubble expectation values. We plan to study this set of equations for the non-Gaussian scaling found in this paper and compare with the large $N$ equations obtained and solved in \cite{SDEs}. Furthermore, all solutions to tensor models are certainly based on this set of equations with different balances, and this is where an interpolation between matrix-like and tensor-like behaviors could be found.

\section*{Acknowledgements}

Research at Perimeter Institute is supported by the Government of Canada through Industry Canada and by the Province of Ontario through the Ministry of Research and Innovation.



\begin{thebibliography}{99}

\bibitem{ambjorn-3d-tensors}
  J.~Ambjorn, B.~Durhuus and T.~Jonsson,
  ``Three-Dimensional Simplicial Quantum Gravity And Generalized Matrix
  Models,''
  Mod.\ Phys.\ Lett.\  A {\bf 6}, 1133 (1991).

\bibitem{sasakura-tensors}
  N.~Sasakura,
  ``Tensor model for gravity and orientability of manifold,''
  Mod.\ Phys.\ Lett.\  A {\bf 6}, 2613 (1991).

\bibitem{gross-tensors}
  M.~Gross,
  ``Tensor models and simplicial quantum gravity in $>$ 2-D,''
  Nucl.\ Phys.\ Proc.\ Suppl.\  {\bf 25A}, 144 (1992).

\bibitem{mm-review-difrancesco}
  P.~Di Francesco, P.~H.~Ginsparg and J.~Zinn-Justin,
  ``2-D Gravity and random matrices,''
  Phys.\ Rept.\  {\bf 254}, 1 (1995)
  [arXiv:hep-th/9306153].

\bibitem{Gur3}
  R.~Gurau,
  ``The 1/N expansion of colored tensor models,''
  Annales Henri Poincare {\bf 12}, 829 (2011)
  [arXiv:1011.2726 [gr-qc]].

\bibitem{GurRiv}
  R.~Gurau and V.~Rivasseau,
  ``The 1/N expansion of colored tensor models in arbitrary dimension,''
  Europhys.\ Lett.\  {\bf 95}, 50004 (2011)
  [arXiv:1101.4182 [gr-qc]].

\bibitem{Gur4}
  R.~Gurau,
  ``The complete 1/N expansion of colored tensor models in arbitrary
  dimension,''
  arXiv:1102.5759 [gr-qc].

\bibitem{1tensor}
  V.~Bonzom, R.~Gurau and V.~Rivasseau,
  ``Random tensor models in the large N limit: Uncoloring the colored tensor models,''
  Phys.\ Rev.\ D {\bf 85}, 084037 (2012)
  [arXiv:1202.3637 [hep-th]].

\bibitem{critical-colored}
  V.~Bonzom, R.~Gurau, A.~Riello and V.~Rivasseau,
  ``Critical behavior of colored tensor models in the large N limit,''
  Nucl. Phys. {\bf B853}, 174-195 (2011).
  [arXiv:1105.3122 [hep-th] ]

\bibitem{harold-hard-dimers}
  V.~Bonzom and H.~Erbin,
  ``Coupling of hard dimers to dynamical lattices via random tensors,''
  J. Stat. Mech. 1209 (2012) P09009.
  arXiv:1204.3798 [cond-mat.stat-mech].

\bibitem{multicritical-dimers}
  V.~Bonzom,
  ``Multicritical tensor models and hard dimers on spherical random lattices,''
  arXiv:1201.1931 [hep-th].

\bibitem{universality}
  R.~Gurau,
  ``Universality for Random Tensors,''
  arXiv:1111.0519 [math.PR].

\bibitem{SDEs}
  V.~Bonzom,
  ``Revisiting random tensor models at large N via the Schwinger-Dyson equations,''
  arXiv:1208.6216 [hep-th].

\bibitem{david-revueDT}
  F.~David,
  ``Simplicial quantum gravity and random lattices,''
  Les Houches Sum. Sch. 1992:0679-750.
  arXiv:hep-th/9303127.

\bibitem{ambjorn-scaling4D}
  J.~Ambjorn and J.~Jurkiewicz,
  ``Scaling in four-dimensional quantum gravity,''
  Nucl.\ Phys.\  B {\bf 451}, 643 (1995)
  [arXiv:hep-th/9503006].

\bibitem{ambjorn-revueDT}
  J.~Ambjorn,
  ``Simplicial Euclidean and Lorentzian quantum gravity,''
  arXiv:gr-qc/0201028.

\bibitem{tree-algebra}
  R.~Gurau,
  ``A generalization of the Virasoro algebra to arbitrary dimensions,''
  Nucl.\ Phys.\  B {\bf 852}, 592 (2011)
  [arXiv:1105.6072 [hep-th]].

\bibitem{bubble-algebra}
  R.~Gurau,
  ``The Schwinger Dyson equations and the algebra of constraints of random tensor models at all orders,''
  Nucl.\ Phys.\ B {\bf 865}, 133 (2012)
  [arXiv:1203.4965 [hep-th]].









































\end{thebibliography}
\end{document}